\documentstyle[12pt]{article}
\newcommand{\be}{\begin{eqnarray}}
\newcommand{\ee}{\end{eqnarray}}

\begin{document}

\vspace{1cm}

\begin{center}

\LARGE{Charge form factor of $\pi$ and $K$ mesons}\\

\vspace{1cm}

\large{ F. Cardarelli$^{(a)}$, I.L. Grach$^{(b)}$, I.M.
Narodetskii$^{(b)}$ \\ E. Pace$^{(a,c)}$, G. Salm\`{e}$^{(d)}$, S.
Simula$^{(d)}$}\\

\vspace{0.5cm}

\normalsize{$^{(a)}$Istituto Nazionale di Fisica Nucleare, Sezione
Tor Vergata\\ Via della Ricerca Scientifica, I-00133 Roma, Italy\\
$^{(b)}$Institute of Theoretical and Experimental Physics\\ Moscow
117259, Russia\\ $^{(c)}$Department of Physics, University of Rome
"Tor Vergata"\\ Via della Ricerca Scientifica, I-00133 Roma, Italy\\
$^{(d)}$Istituto Nazionale  di Fisica Nucleare, Sezione Sanit\`{a},\\
Viale Regina Elena 299, I-00161 Roma, Italy}

\end{center}

\vspace{1cm}

\begin{abstract}

The charge form factor of $\pi$ and $K$ mesons is evaluated adopting a
relativistic constituent quark model based on the light-front
formalism. The relevance of the high-momentum components of the meson
wave function, for values of the momentum transfer accessible to
$CEBAF$ energies, is illustrated. The predictions for the elastic
form factor of $\pi$ and $K$ mesons are compared with the results of
different relativistic approaches, showing that the measurements of
the pion and kaon form factors planned at $CEBAF$ could provide
information for discriminating among various models of the meson
structure.

\end{abstract}

\vspace{1cm}

PACS number(s): 12.39.Ki, 13.40.Gp, 14.40.Aq, 14.40.Lb

\newpage

\pagestyle{plain}

\indent The evaluation of the electromagnetic (e.m.) properties of
$\pi$ and $K$ mesons has recently received a renewed interest,
because measurements of the pion and kaon charge form factors are
planned at $CEBAF$ \cite{PROPOSALS}. In the past few years,
light-front constituent quark models have been extensively applied to
relativistic calculations of various electroweak properties of mesons
\cite{CAR94} - \cite{MESONS} and baryons \cite{NUCLEON}. In most of
these applications (\cite{SCH94} - \cite{NUCLEON}) it is assumed
that: i) the hadron wave function is given by a harmonic oscillator
(HO) ansatz, which is expected to describe the effects of the
confinement scale only, or has a power-law (PL) behaviour, dictated
at large momenta by the perturbative $QCD$ theory \cite{BL80}; ii)
the constituent quarks are point-like objects as far as their e.m.
properties are concerned. In Ref. \cite{CAR94} a different approach
is adopted, namely: ~ i) a light-front mass operator, constructed
from the effective $q \bar{q}$ Hamiltonian of Ref. \cite{GI85}
reproducing the meson mass spectra, is considered and the
corresponding eigenfunctions are used to describe the dynamics of the
constituent quarks inside the meson; ~ ii) a non-vanishing size of
the constituent quarks is assumed and a simple monopole charge form
factor for the constituent quarks is introduced. Within this approach
existing pion data both at low and high values of the squared
four-momentum transfer $Q^2$ are reproduced. Moreover, it has been
shown that the high-momentum components, generated in the wave
function by the one-gluon-exchange (OGE) part of the effective $q
\bar{q}$ interaction of Ref. \cite{GI85}, sharply affect the pion
charge form factor for values of $Q^2$ up to few $(GeV/c)^2$, i.e. in
a range of values accessible to $CEBAF$ energies. Differently, in Ref.
\cite{SCH94} it has been claimed that the charge form factor of
pseudoscalar mesons is insensitive to a large class of wave
functions, and, moreover, that the high-momentum tail of the wave
function does not matter for energies accessible to present
experiments. The aims of this brief report are i) to point out that
our wave functions do not belong to the limited class of wave
functions considered in \cite{SCH94}, and ii) to clarify the
relevance of the high-momentum components of the meson wave function,
particularly for values of $Q^2 \sim$ few $(GeV/c)^2$, by analyzing
in detail the structure of the expression of the pion form factor
used in Refs. \cite{CAR94} and \cite{SCH94}. Moreover, our
theoretical predictions for the elastic form factor of $\pi^+$, $K^+$
and $K^0$ mesons are compared with the results obtained within
different sophisticated relativistic approaches, showing that the
measurements of the pion and kaon form factors planned at $CEBAF$
\cite{PROPOSALS} could provide information for discriminating among
various models of the meson structure.

\indent We will start directly from the general expression of the
charge form factor of a pseudoscalar meson, $F^{PS}(Q^2)$, obtained
within the light-front constituent quark model (see, e.g., Ref.
\cite{CAR94}), viz.
 \be
    F^{PS}(Q^2) = e_q f^q(Q^2) H^{PS}(Q^2; m_q, m_{\bar{q}}) +
    e_{\bar{q}} f^{\bar{q}}(Q^2) H^{PS}(Q^2; m_{\bar{q}}, m_q)
    \label{1}
 \ee
where $e_q$ ($m_q$) is the charge (mass) of the constituent quark and
$f^q(Q^2)$ its charge form factor. In Eq. (\ref{1}) the body form
factor $H^{PS}(Q^2; m_1, m_2)$ is given explicitely by
 \be
    H^{PS}(Q^2; m_1, m_2) = \int d\vec{k}_{\perp}~ d\xi ~ { \sqrt{M_0
    M'_0} \over 4 \xi (1 - \xi) } ~ \sqrt{ \left [ 1 - \left( {m_1^2 -
    m_2^2 \over M_0^2 } \right ) ^2 \right ] \left [ 1 - \left (
    {m_1^2 - m_2^2 \over {M'}_0^2} \right ) ^2 \right ] }
    \nonumber \\
    \cdot ~ {w^{PS}(k^2) w^{PS}({k'}^2) \over 4 \pi } ~ { \xi (1 -
    \xi) ~ [M_0^2 - (m_1 - m_2)^2 ] + \vec{k}_{\perp} \cdot
    (\vec{k'}_{\perp} - \vec{k}_{\perp}) \over \xi (1 - \xi)
    \sqrt{M_0^2 - (m_1 - m_2)^2} \sqrt{{M'}_0^2 - (m_1 - m_2)^2} }
    \label{2}
 \ee
where the free mass operator $M_0$ ($M'_0$) and the intrinsic
light-front variables $\vec{k}_{\perp}$ ($\vec{k'}_{\perp}$ ), $\xi$
are defined as
 \be
    M_0^2 = {m_1^2 + k^2_{\perp} \over \xi} + {m_2^2 + k^2_{\perp}
    \over(1 - \xi)} ~~ & , & ~~ {M'}_0^2 = {m_1^2 + {k'}_{\perp}^2
    \over \xi} + {m_2^2 + {k'}_{\perp}^2 \over (1 - \xi)}
    \nonumber \\
    \vec{k}_{\perp} = \vec{p}_{1 \perp} - \xi \vec{P}_{\perp} = -
    \vec{p}_{2 \perp} + (1 - \xi) \vec{P}_{\perp} ~~ & , & ~~
    \vec{k'}_{\perp}  \equiv \vec{k}_{\perp} + (1 - \xi)
    \vec{Q}_{\perp}
    \nonumber \\
    \xi = p_1^+ / P^+ = 1 - p_2^+ / P^+ ~~ & &
    \label{3}
 \ee
In Eqs. (\ref{2}) - (\ref{3}) the subscript $\perp$ indicates the
projection perpendicular to the spin quantization axis, defined by
the vector $\hat{n} = (0,0,1)$, and the {\em plus} component of a
four-vector $p \equiv (p^0, \vec{p} )$ is given by $p^+ = p^0 +
\hat{n} \cdot \vec{p}$. Moreover, $\tilde{P} \equiv (P^+,
\vec{P}_{\perp}) = \tilde{p}_1 + \tilde{p}_2$ is the light-front
momentum of the meson, $k^2 \equiv k_{\perp}^2 + k_n^2$,  ${k'}^2
\equiv {k'}_{\perp}^2 + {k'}_n^2$, $k_n \equiv (\xi - 1 / 2) M_0 +
(m_2^2 - m_1^2) / 2 M_0$ and $k'_n \equiv (\xi - 1 / 2) M'_0 + (m_2^2
- m_1^2) / 2 M'_0$.

\indent Following Ref. \cite{CAR94}, the radial wave function
$w^{PS}(k^2)$ appearing in Eq. (\ref{2}) can be identified with the
equal-time radial wave function in the meson rest-frame. In what
follows, we will make use of the eigenfunctions of the effective $q
\bar{q}$ Hamiltonian, developed by Godfrey and Isgur ($GI$)
\cite{GI85} to reproduce the meson mass spectra. In case of
pseudoscalar mesons one has
 \be
    H_{q \bar{q}} ~ w^{PS}(k^2) | 0 0 \rangle \equiv \left
    [\sqrt{m_q^2 + k^2} + \sqrt{m_{\bar{q}}^2 + k^2} + V_{q \bar{q}}
    \right ] ~ w^{PS}(k^2) | 0 0 \rangle  =  M_{q \bar{q}} w^{PS}(k^2)
    | 0 0 \rangle
    \label{5}
 \ee
where $M_{q \bar{q}}$ is the mass of the meson, $| 0 0 \rangle =
\sum_{\nu \bar{\nu}} \langle {1 \over 2} \nu {1 \over 2} \bar{\nu} |
0 0 \rangle \chi_{\nu} \chi_{\bar{\nu}}$ is the usual quark-spin wave
function of a pseudoscalar meson and $V_{q \bar{q}}$ is the effective
$q \bar{q}$ potential. The $GI$ interaction, $V_{(GI)}$, is composed
by a $OGE$ term (dominant at short separations) and a
linear-confining term (dominant at large separations). We will
consider two types of wave functions: the first one is given by the
solution of Eq. (\ref{5}) obtained when the $OGE$ part of $V_{(GI)}$
is switched off, i.e., when only its linear confining term,
$V_{(conf)}$, is retained, whereas the second choice is obtained by
solving Eq. (\ref{5}) with the full $GI$ interaction. The two
different forms of $w^{PS}(k^2)$ will be denoted hereafter by
$w^{PS}_{(conf)}$ and $w^{PS}_{(GI)}$ corresponding to $V_{(conf)}$
and $V_{(GI)}$, respectively. Note that the pion mass corresponding
to $V_{(conf)}$ in Eq. (\ref{5}) is $1.024 ~ GeV$, whereas the one
obtained using $V_{(GI)}$ is $0.149 ~ GeV$. The pion wave functions
$w_{(conf)}^{\pi}$ and $w_{(GI)}^{\pi}$ are shown in Fig. 1 and
compared with the HO ($w_{(HO)}^{\pi} \propto exp(- k^2 / 2
\alpha^2)$) and PL ($w_{(PL)}^{\pi} \propto (1 + k^2 /
\beta^2)^{-2}$) wave functions used in Ref. \cite{SCH94}. It should
be stressed that the latter ones are constrained by imposing
the reproduction of the leptonic decay constants of $\pi$ and $\rho$
mesons and by assuming a point-like quark electroweak (e.w.) current.
It can clearly be seen that: i) the momentum behaviours of
$w_{(conf)}^{\pi}$ and $w_{(GI)}^{\pi}$ are sharply different,
because of the configuration mixing induced by the $OGE$ part of the
effective $q \bar{q}$ interaction; ii) for $k < 1 ~ GeV/c$
$w_{(HO)}^{\pi}$ and $w_{(PL)}^{\pi}$ are quite similar (possibly
because they have to fulfil the above-mentioned constraints) and do
not differ significantly from $w_{(conf)}^{\pi}$, which takes into
account the effects of the confinement scale only; iii) the
high-momentum tail of $w_{(GI)}^{\pi}$, while exhibiting a nominal
power-law fall off at large momenta, is much higher than the one
pertaining to $w_{(PL)}^{\pi}$. The average transverse momentum
$\bar{k}_{\perp} \equiv \sqrt{<k_{\perp}^2>}$ turns out to be $\simeq
0.8 ~ GeV/c$ in case of $w_{(GI)}^{\pi}$ and $\simeq 0.3 ~ GeV/c$ for
$w_{(HO)}^{\pi}$, $w_{(PL)}^{\pi}$ and $w_{(conf)}^{\pi}$. Thus, the
HO and PL wave functions adopted in Refs. \cite{SCH94} and
\cite{NUCLEON}(c) can hardly be considered representative of the
range of uncertainty of the momentum behaviour of the wave function.
As a matter of fact, our $w_{(GI)}^{\pi}$ wave function, which is
eigenfunction of a mass operator reproducing the meson mass spectra,
does not belong to the limited class of wave functions considered
in Refs. \cite{SCH94} and \cite{NUCLEON}(c), since it gives rise to an
overestimation of the leptonic decay constants when a point-like
quark e.w. current is adopted (cf. \cite{CAR94}).

\indent The relevance of the high-momentum components of the wave
function in the calculation of the pion form factor can be
investigated by considering in Eq. (\ref{2}) different values of the
upper limit of integration over $k_{\perp} \equiv |\vec{k}_{\perp}|$
(denoted hereafter by $k_{\perp}^U$). The results of the
calculations, obtained assuming $f^q=1$ in Eq. (\ref{1}) and using in
Eq. (\ref{2}) both $w_{(conf)}^{\pi}$ and $w_{(GI)}^{\pi}$, are shown
in Fig. 2 for values of $Q^2$ up to $10~(GeV/c)^2$. In what follows
we will limit ourselves to consider the wave functions
$w_{(conf)}^{\pi}$ and $w_{(GI)}^{\pi}$, because for $Q^2 <
10~(GeV/c)^2$ the results obtained using the HO and PL wave functions
of Ref. \cite{SCH94} do not differ significantly from those
calculated with $w_{(conf)}^{\pi}$. From Fig. 2 it can clearly be
seen that, both for $w_{(conf)}^{\pi}$ and $w_{(GI)}^{\pi}$, the
calculation of the pion charge form factor is strongly affected by
components of the wave function corresponding to $k_{\perp} ~ > ~
\bar{k}_{\perp}$. As a matter of fact, in case of $w_{(conf)}^{\pi}$,
$\sim 90 \%$ of the form factor at $Q^2 > 0.5 ~ (GeV/c)^2$ is due to
components of the wave functions with $k_{\perp} > 0.3~GeV/c$
($\simeq ~(\bar{k}_{\perp})_{conf}$), and, moreover, the saturation
is almost reached only when $k_{\perp}^U \simeq 1.5 ~ GeV/c$ ($\sim 5
{}~ (\bar{k}_{\perp})_{conf}$). In case of $w_{(GI)}^{\pi}$, the
high-momentum tail corresponding to $k_{\perp} > 0.8 ~ GeV/c$ ($\simeq
{}~ (\bar{k}_{\perp})_{GI}$) is responsible for $\sim 50 \%$ of the
pion form factor at $Q^2 > 0.5 ~ (GeV/c)^2$ and the saturation at high
values of $Q^2$ is almost reached only when $k_{\perp}^U \simeq
2.5~GeV/c$ ($\simeq 3 ~ (\bar{k}_{\perp})_{GI}$). Such results are
simply related to the fact that, for $Q \sim$ few $GeV/c$,  values of
$k_{\perp} \sim 1 ~ GeV/c$ can give rise to low values of $k'_{\perp}$
($= |\vec{k}_{\perp} + (1 - \xi) \vec{Q}_{\perp}|$), when
$\vec{k}_{\perp}$ is antiparallel to $\vec{Q}_{\perp}$ and the struck
quark carries an average fraction of the momentum of the meson (i.e.,
$\xi \sim \bar{\xi} = 0.5$ in the pion) \footnote[1]{Note that the
$\xi$-distribution corresponding both to $w_{(conf)}^{\pi}$ and
$w_{(GI)}^{\pi}$ exhibits a flat maximum in the region $0.2 < \xi <
0.8$}. This means that for $Q \sim$ few $GeV/c$ configurations both
at short and large transverse $q \bar{q}$ separations are relevant
(see the product $w^{PS}(k^2) w^{PS}({k'}^2)$ in the integrand of
Eq.(\ref{2})). To sum up, the results reported show that: i) the
momentum behaviour of the wave function at $k > 1 ~ (GeV/c)$ can play
a relevant role in determining the pion form factor for values of
$Q^2$ accessible to $CEBAF$ energies; ii) according to the findings
of Ref. \cite{CAR94} the pion form factor is sharply overestimated
due to the effects of the high-momentum components generated in the
wave function by the $OGE$ part of the $GI$ interaction (compare solid
lines in Fig. 2(a) and 2(b)), which, as known, nicely explains the
$\pi$ - $\rho$ mass splitting. In this work we have checked that the
same conclusions hold as well for the charge form factor of $K$ meson,
whereas they are no longer true in case of heavy pseudoscalar mesons,
like the $D$ and $B$ mesons. As a matter of fact, the explicit
calculations of Eqs. (\ref{1}-\ref{2}) (assuming $f^q=1$) yield
almost the same results in a wide range of values of $Q^2$ ($Q^2 >>
1~(GeV/c)^2$) both for $w_{(conf)}^{D(B)}$ and $w_{(GI)}^{D(B)}$ wave
functions. We will limit ourselves to comment that such a result can
be ascribed to the fact that: ~ i) the body form factor ($H^{PS}$)
corresponding to the virtual photon absorption by the heavy $c$ ($b$)
quark in the $D$ ($B$) meson is dominant; ~ ii) the average fraction
of the momentum of the meson carried by the heavy quark is very close
to $1$, leading to $k'_{\perp} \simeq k_{\perp}$, which implies a
weak dependence of the calculated form factor on the heavy meson wave
function in a wide range of values of $Q^2$.

\indent The results reported in Fig. 2 suggest that, if the
constituent quarks are assumed to be point-like particles (i.e., if
$f^q=1$), the pion form factor calculated with wave functions having
$\bar{k}_{\perp} \sim 0.3 ~ GeV/c$ (like, e.g., $w_{(conf)}^{\pi}$,
$w_{(HO)}^{\pi}$ and $w_{(PL)}^{\pi}$) is in fairly good agreement
with existing data, whereas the one obtained using $w_{(GI)}^{\pi}$
is not. However, once the assumption $f^q=1$ is made, the parameters
which unavoidably appear in the hadron wave function are usually
adjusted in order to fit e.m. (or, more generally, electroweak)
hadron properties (see, e.g., Ref. \cite{NUCLEON}(c)). In this way
the relativistic constituent quark model (RCQM) looses (at least
partially) its predictive power, for the wave function is not
completely independent of the e.m. observable under investigation. A
different approach is to adopt the eigenfunctions of a (light-front)
mass operator able to reproduce correctly the hadron mass spectra, so
that the hadron wave functions do not depend upon any observable but
the hadron energy levels. In this way, the momentum behaviour
of the hadron wave functions is dictated by the features of the
effective $q \bar{q}$ interaction appearing in the mass operator and
the investigation of the e.m. properties of hadrons could provide
information on those of the constituent quarks. Thus, in order to
recover the predictive power of the $RCQM$, the same e.m. one-body
current should be used for all the hadrons. Following this strategy,
a simple monopole ansatz for the charge form factor of the
constituent u and d quarks has been considered in Ref.
\cite{CAR94}, viz.
 \be
    f^q(Q^2) = {1 \over 1 + Q^2 <r^2>_q / 6}
    \label{6}
 \ee
When the wave function $w_{(GI)}^{\pi}$ is adopted in Eq. (\ref{2}),
the value $<r^2>_u = <r^2>_d = (0.48 ~ fm)^2$ has to be chosen in
order to reproduce the experimental value of the pion charge radius
$<r^2>_{exp}^{(\pi)} = (0.660 ~ \pm ~ 0.024 ~ fm)^2$. It should be
pointed out that such a value of the constituent quark radius is in
nice agreement with the ansatz $<r^2>_q = \kappa / m_q^2$, suggested
in Ref. \cite{PH90} from the analysis of the so-called strong
interaction radius of hadrons, when the values $\kappa \simeq 0.3$,
extracted from the chiral quark model of Ref. \cite{VLKW90}, and $m_u
= m_d = 0.220 ~ GeV$ \cite{GI85} are adopted \footnote[2]{Note that
the difference between the $\rho$ and $\pi$ radii found in Ref.
\cite{PH90} (i.e., $<r^2>^{(\rho)} - <r^2>^{(\pi)} = 0.11 \pm 0.06 ~
fm^2$) is independent of the constituent quark radius and is nicely
explained by the configuration mixing due to the spin-dependent part
of the effective $q \bar{q}$ interaction, as it can be inferred from
the results of Refs. \cite{CAR94,CAR95} yielding $<r^2>^{(\rho)} -
<r^2>^{(\pi)} = 0.14 ~ fm^2$.}. Moreover, it should be stressed that,
though the u(d)-quark charge radius is fixed only by the pion data at
very low values of $Q^2$, the predictions of our RCQM compare very
favourably with the data also at high values of $Q^2$ (see Ref.
\cite{CAR94}). This is illustrated in Fig. 3, where  our results for
the pion charge form factor are compared with the experimental data
\cite{PION-DATA} and also with the predictions of different
sophisticated relativistic approaches, like the covariant
Bethe-Salpeter approach of Ref. \cite{BS} and the QCD sum rule
technique of Ref. \cite{NR82}. The predictions of the simple Vector
Meson Dominance (VMD) model, including the $\rho$ - meson pole only,
are also shown in the same figure. It can be seen that existing pion
data do not discriminate among calculations based on different models
of the pion structure.

\indent  By using in Eq. (\ref{2}) the appropriate eigenfunctions of
the $GI$ Hamiltonian (\ref{5}), the elastic form factors of charged
$K^+$ and neutral $K^0$ mesons have been calculated. In Fig. 4 the
results of our calculations, performed adopting different choices of
the charge radius of the constituent s quark ($<r^2>_s$), are
reported and compared with the predictions of Ref. \cite{BS}, based
on a covariant Bethe-Salpeter approach. It can be seen that for $Q^2
> 1 ~ (GeV/c)^2$ the calculated charge form factors of $K^+$ and $K^0$
mesons are remarkably sensitive to the value used for $<r^2>_s$, so
that their experimental investigation could provide information on
the e.m. structure of light constituent quarks. From Fig. 4 it can
also be seen that, unlike the case of the pion, the measurement of
the kaon form factor at $Q^2 > 1 ~ (GeV/c)^2$ could discriminate among
different models of the meson structure.

\indent In conclusion, the charge form factor of $\pi$ and $K$ mesons
has been evaluated within a light-front constituent quark model. The
use of the eigenfunctions of a mass operator, constructed from the
effective $q \bar{q}$ Hamiltonian of Ref. \cite{GI85} reproducing the
meson mass spectra, and the introduction of a phenomenological charge
form factor for the constituent quarks have been briefly discussed. It
has been shown that the high-momentum tail of the meson wave function
(namely, $k > 1 ~ GeV/c$) is essential in determining the behaviour of
the form factor already at $Q^2 > 0.5 ~ (GeV/c)^2$. Thus, the
investigation of $\pi$ and $K$ form factors at $CEBAF$ represents a
powerful tool to study the short-range structure of mesons. The
predictions of our relativistic constituent quark model for the
charge form factor of $\pi$ and $K$ mesons have been compared with
those of different sophisticated relativistic approaches, showing
that the planned experiments at $CEBAF$ \cite{PROPOSALS}, aimed at
measuring independently the pion and kaon form factor for $Q^2 <
3~(GeV/c)^2$, could provide relevant information on the
electromagnetic structure of light constituent quarks and could
represent an interesting tool to discriminate among different models
of the meson structure.

\vspace{0.25cm}

\indent We gratefully acknowledge S. Brodsky for helpful discussions
and R.A. Williams for supplying us with the numerical output of the
$K^+$ and $K^0$ calculations of Ref. \cite{BS}.

\newpage

\vspace{0.5cm}

\begin{center}

{\bf Figure Captions}

\end{center}

\vspace{0.5cm}

Fig. 1. Pion wave functions $(k \cdot w^{\pi}(k^2))^2$, calculated
using in Eq. (\ref{5}) different effective $q \bar{q}$ interactions,
as a function of the relative momentum $k$. Dotted line:
$w_{(conf)}^{\pi}$, corresponding to the case in which only the
linear confining part of the $GI$ $q \bar{q}$ interaction \cite{GI85}
is considered. Solid line: $w_{(GI)}^{\pi}$, corresponding to the
solution of Eq. (\ref{5}) obtained using the full $GI$ $q \bar{q}$
interaction. The dot-dashed and dashed lines correspond to the
harmonic oscillator ($w_{(HO)}^{\pi} \propto exp(- k^2 / 2
\alpha^2)$) and power-law ($w_{(PL)}^{\pi} \propto (1 + k^2 /
\beta^2)^{-2}$) wave functions introduced in Ref. \cite{SCH94},
respectively.

\vspace{0.5cm}

Fig. 2. Charge form factor of the pion, $Q^2 ~ F^{\pi}(Q^2)$,
calculated assuming $f^q=1$ in Eq. (\ref{1}) and using in Eq.
(\ref{2}) the wave functions $w_{(conf)}^{\pi}$ (a) and
$w_{(GI)}^{\pi}$(b). The various lines correspond to the results
obtained assuming in Eq. (\ref{2}) different values of $k_{\perp}^U$,
the upper limit of integration over $|\vec{k}_{\perp}|$. The dotted
and dashed lines correspond to $k_{\perp}^U = 0.3$ and $0.8~GeV/c$,
respectively, whereas the dot-dashed lines correspond to $k_{\perp}^U
= 1.5~GeV/c$ in (a) and $2.5~GeV/c$ in (b). The solid lines represent
the full calculations of the elastic form factor (i.e., when
$k_{\perp}^U \rightarrow \infty$). The experimental data are taken
from Ref. \cite{PION-DATA}.

\vspace{0.5cm}

Fig. 3. Elastic form factor of the charged pion, times $Q^2$, as a
function of $Q^2$. The solid line represents the results of our
relativistic constituent quark model (RCQM), obtained using in Eq.
(\ref{2}) the appropriate eigenfunction of the effective $q \bar{q}$
Hamiltonian of Ref. \cite{GI85} (see Eq. (\ref{5})) and adopting in
Eq. (\ref{1}) the monopole charge form factor (Eq. (\ref{6}) with the
quark charge radius equal to $<r^2>_u = <r^2>_d = (0.48~fm)^2$. The
dashed and dot-dashed lines represent the predictions of the
covariant Bethe-Salpeter approach of Ref. \cite{BS} and of the QCD
sum rule technique of Ref. \cite{NR82}, respectively. The dotted line
is the prediction of a simple VMD model, which includes the $\rho$ -
meson pole only (i.e., $F^{\pi} = (1 + Q^2 / m_{\rho}^2)^{-1}$).

\vspace{0.5cm}

Fig. 4. Elastic form factor of charged $K^+$ (a) and neutral $K^0$
(b) mesons, times $Q^2$, as a function  of $Q^2$. The solid line
represents the results of our relativistic constituent quark model
(RCQM), obtained using in Eq. (\ref{2}) the appropriate
eigenfunctions of the effective $q \bar{q}$ Hamiltonian of Ref.
\cite{GI85} (see Eq. (\ref{5})) and adopting in Eq. (\ref{1}) a
$SU(3)$ symmetric (monopole) charge form factor for the constituent
quarks (i.e., $f^u = f^d = f^s$) with the charge radius $<r^2>_u =
<r^2>_d = <r^2>_s = (0.48~fm)^2$. The dot - dashed lines are the
results of the calculations of Ref. \cite{BS}, based on a covariant
Bethe-Salpeter approach. The dashed lines represent the predictions
of our RCQM, calculated using different values for the charge radius
of the constituent s and u(d) quarks, namely $<r^2>_s = (0.25~fm)^2$
and $<r^2>_u = <r^2>_d = (0.48~fm)^2$. Note that these values
correspond to the ansatz $<r^2>_q = \kappa / m_q^2$ \cite{PH90},
adopting $\kappa \simeq 0.3$ \cite{VLKW90}, $m_u = m_d = 0.220 ~ GeV$
and $m_s = 0.419 ~ GeV$ \cite{GI85}. Eventually, the dotted line in
(a) is the prediction of the VMD model including the $\rho$ - meson
pole only.

\end{document}